\documentclass[12pt]{article}
\usepackage{epsfig}

\newcommand{\gapprox}{{\raisebox{-0.5ex}{${\scriptstyle>}$} \atop \raisebox{0.5ex}{${\scriptstyle\sim}$}}}
\newcommand{\lapprox}{{\raisebox{-0.5ex}{${\scriptstyle<}$} \atop \raisebox{0.5ex}{${\scriptstyle\sim}$}}}

\title{The Influence of Medium Effects on the Gross Structure of Hybrid Stars
\footnote{Supported by BMBF, GSI Darmstadt, and DFG.}\addtocounter{footnote}{5}
$^,$\footnote{This paper forms part of the dissertation of K. Schertler.}
}
\author{K.~Schertler\addtocounter{footnote}{-4}\footnote{E-mail: klaus.schertler@theo.physik.uni-giessen.de}, C.~Greiner, 
P.K.~Sahu\addtocounter{footnote}{-3}\footnote{Alexander von Humboldt Research Fellow.}, and M.H.~Thoma\addtocounter{footnote}{2}\footnote{Heisenberg Fellow.}}
\date{}
\begin{document}
\maketitle
\begin{center}
{\small \it Institut f\"ur Theoretische Physik, Universit\"at Giessen \\
35392 Giessen, Germany }
\end{center}
%
%
\begin{figure}[ht]
\centerline{\epsfig{file=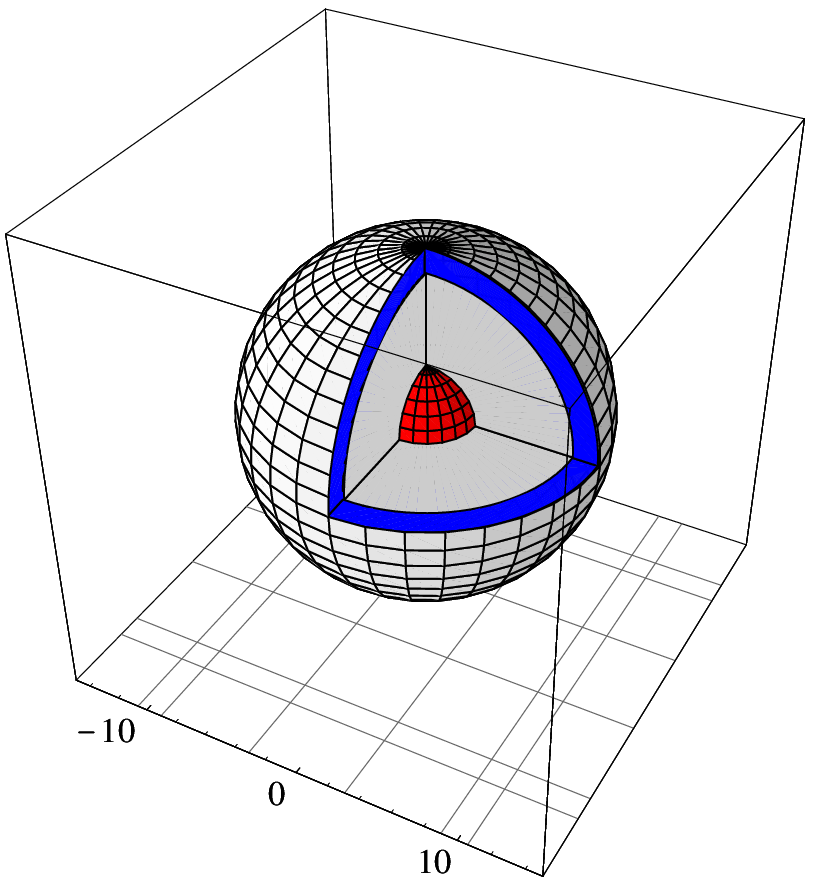,height=4cm}}
\end{figure}
%
\begin{abstract}
We investigate the influence of medium effects on the structure 
of hybrid stars, i.e.~neutron stars possessing a quark matter core.  
We found that medium effects in quark matter reduce the extent of a pure quark matter phase 
in the interior of a hybrid star significantly in favor of a mixed phase
of quark and hadronic matter. Over a wide range of the strong coupling constant
-- which parameterizes the influence of medium effects -- quark matter is able to
exist at least in a mixed phase in the interior of neutron stars.


\end{abstract}

\section{Introduction}

The composition and properties of matter under extreme conditions as they can be found
in the deep dense interior of neutron stars has been a subject of investigation over the last
six decades. The large densities inside the star should be sufficient to bear 
new ``exotic'' strange components like hyperons \cite{Glen8287, SchaMish96}, kaons \cite{LiLeeBrow97, SchaGlen98} 
or even to melt the baryons to form a deconfined quark matter phase. 
The appearance of such components may significantly change the gross structure 
of the star like its mass and radius.

Since the pioneering work two decades ago \cite{BaymChin76, ChapNaue76, KeisKiss76, FreeMcLe78a}
the fascinating possibility of a deconfined quark matter phase in the 
interior of neutron stars has stimulated the work of many authors 
\cite{Glen92, HeisPethStau9394, GlenPei95, Prak, HybRecent, GlenBook}. Moreover, recently Glendenning, 
Pei and Weber \cite{GlenPeiWebe97} proposed that an observable signature in the timing 
structure of pulsars could even provide a signal for the existence of such a quark matter phase in 
the cores of pulsars. Neutron stars which are made of hadronic matter in the outer region, 
but possessing a strange quark matter (SQM) core in their interior are called hybrid stars.
The essential uncertainty for all theoretical investigations dealing with the question whether or
not such phase exists, is clearly the equation of state (EOS) from about normal nuclear matter density $\rho_0$
to the order of $\sim 10\,\rho_0$ achieved in the center of the star. 
So far, there is unfortunately no single theory to cover this
density range with respect to quark degrees of freedom. We therefore still depend on separate 
descriptions including the basic degrees of freedom of dense hadronic matter and quark matter,  
respectively. 
The EOS for the description of SQM in the framework of the commonly used
MIT bag model was recently improved by including medium effects \cite{Sche97}. 
This was done by using a quasi-particle approach, successfully used in various parts of physics.
The interaction of the quarks with the other quarks of the system is implemented by giving them
density dependent effective quark masses. 
This should lead to a more realistic description 
of quark matter going beyond the free Fermi gas approximation.
We will refer to this model \cite{Sche97} as the ``effective mass bag model''.
It was found that medium effects increase the energy per baryon of SQM
and therefore make the SQM phase energetically less favorable.
The physical reason for the larger energy per baryon is 
the increase of the effective quark masses in medium \cite{Sche97}.

The scope of this work is to study the influence of these medium effects on the gross structure 
of non-rotating hybrid stars and on the existence of the SQM phase inside the star. 
In section \ref{EffectMassBagModel}, we review the inclusion of medium effects 
in the EOS of SQM within the effective mass bag model \cite{Sche97}. 
In section \ref{HadronicEOS}, we shortly discuss the used hadronic EOS, before we come
to the construction of the phase transition between quark and hadronic matter in
section \ref{Construction}. 
Finally, we discuss in detail the results of the influence of medium effects on the gross structure of hybrid stars 
in section \ref{GrossStructure}. The conclusions are presented in section \ref{conclusion}.

\section{Quark matter in the effective mass bag model} \label{EffectMassBagModel}
 
\subsection{The model}

The strange quark matter phase containing an approximately equal amount of $u$, $d$ and $s$-quarks
(SQM) has been suggested as a possible stable or metastable phase of nuclear matter \cite{Bodm71Witte84}.
The EOS of SQM is usually described in the framework of a bag model i.e.~as 
a non-interacting Fermi gas of quarks at zero temperature, 
taking into account the bag constant \cite{Bodm71Witte84,FahrJaff84}. 
Quark interactions within lowest order 
perturbative QCD have also been considered \mbox{\cite{FahrJaff84,FreeMcLe78b}}.

In all parts of physics from solid-state physics to nuclear and high-energy physics \cite{Lattice},
the quasi particle picture is successfully used to describe complicated interactions
and collective effects like medium effects in many particle systems.
The particles of such systems are thought to acquire an effective mass by the 
interaction of the particles with the system.
Adopting this quasi particle approach, recently medium effects were included 
in the EOS of SQM in the framework of the MIT bag model \cite{Sche97}.
There, the quarks are considered as quasi-particles which acquire an effective 
mass by the interaction with the other quarks of the dense system. 
%
%
\begin{figure}[hb]
\centerline{\epsfig{file=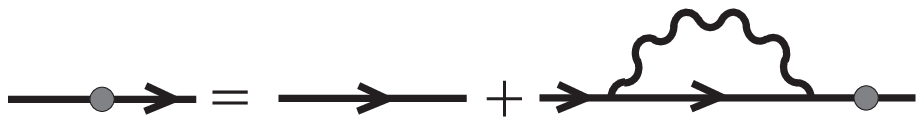,height=1cm}}
\caption{Resummed quark propagator.}
\label{dyson}
\end{figure}
%
The effective quark masses follow from the pole of the resummed one-loop quark propagator at finite chemical potential
(fig.~\ref{dyson}) which is calculated in the hard dense loop approximation \cite{Sche97, EffeMass}.
One finds
\begin{equation} \label{defmq}
{m_q^*}(\mu)=\frac{g \mu}{\sqrt{6}\pi}
\end{equation}
for the light quarks (i.e.~$u$,$d$-quarks) and
\begin{equation} \label{defms}
m_s^*(\mu)=\frac{m_s}{2}+\sqrt {\frac{m_s^2}{4}+\frac{g^2\mu ^2}{6\pi ^2}}
\end{equation}
for massive quarks with current quark mass $m_s$ ($s$-quarks). Note that the effective masses
$m_q^*$ and $m_s^*$
increase with the strong coupling constant $g$ and the quark chemical potential $\mu$.
The effective masses (\ref{defmq}), (\ref{defms}) are used in the ideal Fermi gas EOS at 
temperature $T=0$. The thermodynamic self-consistency of the EOS is fulfilled in the 
sense of \cite{GoreYang95}. Finally this corresponds to inserting the effective masses 
$m^*(\mu)$ directly into the Fermi gas expression for the one particle density 
(where $d$ is the degree of degeneracy)
\begin{equation}
  \rho(\mu) = \frac{d}{6 \pi^2} (\mu^2-{m^*}^2(\mu))^{3/2},
\end{equation}
and extracting the pressure $p(\mu)$ and energy density 
$\epsilon(\mu)$ from the thermodynamic relations
\begin{equation}
        \rho(\mu) = \frac{d p(\mu)}{d \mu} \quad \mbox{and} \quad \epsilon(\mu)+p(\mu) = \mu \rho(\mu).
\end{equation}
For more details see \cite{Sche97}.
The overall pressure $p_{QP}$ and energy density $\epsilon_{QP}$ of the quark phase (QP) is the sum of the expressions for $u$,
$d$ and $s$-quarks plus the corresponding Fermi gas expressions for the uniform background of electrons.
A phenomenological bag constant $B$ is introduced in the usual way by adding it to the energy density and subtracting 
it from the pressure
\begin{eqnarray} \label{bag}
\epsilon_{QP} & = & \epsilon_u+\epsilon_d+\epsilon_s+\epsilon_e +B, \\
p_{QP} & = & p_u +p_d +p_s +p_e -B. \nonumber
\end{eqnarray}
The bag constant $B$ is supposed to mimic the influence of confinement and
corresponds to the energy difference between the perturbative vacuum inside
the deconfined quark matter phase and the ``true'' vacuum outside.

It was found in \cite{Sche97} that the energy per
baryon of SQM increases with increasing coupling constant $g$ due to an increase of the effective masses 
(\ref{defmq}), (\ref{defms}). 
This makes the QP energetically less favorable which -- as we
will see -- will reduce the extent of a pure QP inside the hybrid star.


\subsection{The model parameters}

Now we want to discuss the parameters that enter into the description of quark matter in the effective mass 
bag model. The most important parameters are the bag constant $B$ and the coupling constant $g$.
For the current quark masses we assume $m_u=m_d=0$ for $u$ and $d$-quarks and \mbox{$m_s=150$ MeV} for $s$-quarks.

\subsubsection*{Bag constant}

We have already mentioned that the phenomenological bag constant $B$ is directly related to 
the pressure $p_{QP}$ which finally determines which kind of phase is realized at given chemical potential
(we will discuss this construction in detail in section \ref{Construction}).
The choice of its value has a strong impact on the gross structure of the star.
Unfortunately, the exact value of the bag constant describing a quark system is far from being settled. This is especially
true in connection with a quasi particle model like ours which may lead to a smaller effective $B$ 
as in the case of a free model \cite{Lattice}.
We therefore take $B$ as a rather free parameter and choose a value which is appropriate for our purpose to investigate the
influence of medium effects.
Depending on the corresponding hadronic EOS, a too low $B$ (e.g.~\mbox{$B^{1/4} \lapprox 140$ MeV})
would prevent the QP (medium effects neglected) from being converted to hadronic matter, 
while a too large value 
\mbox{($B^{1/4} \gapprox 200$ MeV)} would lead to a pure hadronic star. Since in this work we are mainly interested 
in the influence of medium effects on hybrid stars, we choose a bag constant of 
\[B^{1/4}\!=165\,\mbox{MeV} \quad \mbox{(}B\approx96\,\mbox{MeV/fm}^3\mbox{).} \] 
Indeed, if we neglect medium effects, we will see that this value ensures the existence of a pure quark core 
while the star is surrounded by a pure hadronic region. 
Nevertheless, we stress that different results e.g.~for 
the absolute values of the quark core radius can be obtained by varying the bag constant \cite{Prak}. Such 
variations will be systematically studied in a future work \cite{ScheWIP}.

As a final remark, we want to note that for our choice of the bag constant  
SQM in equilibrium ($p_{QP}=0$) is not absolutely stable with respect to hadronic matter as it was 
supposed by Witten \cite{Bodm71Witte84}. Neglecting medium effects, we obtain an energy per baryon of
\[\left(\frac{E}{A}\right)_{QP} = \, 3 \cdot \frac{\epsilon_{QP}}{\rho_{QP}} \! \approx 990\, \mbox{MeV,}  \]
which is clearly above the energy per nucleon in $^{56}$Fe.
A larger bag constant or the inclusion of medium effects will further increase $(E/A)_{QP}$ (see \cite{Sche97}, fig. 5).

\subsubsection*{Coupling constant}

The coupling constant $g$ which enters into the effective quark masses (\ref{defmq}) and (\ref{defms}) is
treated in two different ways:
\begin{enumerate}
\item Fixed coupling constant $g=$ const. \quad (sect.~\ref{gconst})
\item Running coupling constant $g=g(\mu)$. \quad (sect.~\ref{grunning})
\end{enumerate}
The first way is thought to study the influence of medium effects parameterized by a constant $g$. Since
at $g=0$ the quarks possess their current quark masses which corresponds to the free EOS (no medium effects),
we can enlarge the influence of medium effects by increasing the value of $g$. The second way should include
the behavior of asymptotic freedom by virtue of a density dependent running coupling constant.

In both cases we have to take the perturbative nature of the effective masses (\ref{defmq}) and (\ref{defms})  
into consideration. Therefore, we restrict ourselves -- for some convenience -- to coupling constants 
$g\lapprox4$ ($\alpha_s \lapprox 1$). 
As we will see, values of $g$ in this range do have a sizeable effect on the EOS and therefore on the structure
of hybrid stars.
Even at smaller $g$ ($g\lapprox2$, $\alpha_s \lapprox 0.3$) modifications to $g=0$ will be realized.
The applicability of results obtained in the hard dense loop limit (see section \ref{EffectMassBagModel}) 
with $g$ taken in this range is discussed in \cite{PeshScheThom97} for the similar hard thermal loop limit. 
It was found that the use of hard thermal loop effective masses is justified from the following point of view. 
The deviations of the hard thermal loop masses from the complete one-loop effective masses (obtained without using the 
hard thermal loop approximation) are indeed small and thus can be neglected in this kind of phenomenological models.


\section{The hadronic EOS\label{HadronicEOS}} 

To describe the hadronic phase of hybrid stars, we use an EOS calculated in the framework of the nonlinear Walecka 
model including nucleons, hyperons ($\Lambda$ and $\Sigma^-$), electrons, muons and $\sigma$, $\omega$ and $\rho$ mesons.
Since the first inclusion of the full octet of the lowest lying baryons by Glendenning \cite{Glen8287}, such relativistic 
mean-field models are widely used for the description of dense nuclear matter \cite{SchaMish96, GlenBook, RMF}.  

Since it is not the purpose of this work to study the influence of different hadronic models, we use a typical hadronic
EOS as an input and refer to \cite{GlenBook} for a
general introduction into relativistic mean-field models and to \cite{Gosh95} for a detailed discussion of our particular EOS and the 
underlying coupling constants. We choose a compression modulus of \mbox{$K=300$ MeV}. The corresponding
hadronic EOS up to two times normal nuclear density will be shown later (fig.~\ref{eos}a, most upper curve).
For subnuclear densities we use the Baym-Pethick-Sutherland (BPS) EOS \cite{GlenBook, BPS} joined to the hadronic EOS.
The BPS EOS mainly influences the low-mass behavior of the mass-radius relation. It leads to an increasing 
radius with a decreasing mass, which is typically for a gravitationally bound star \cite{GlenBook}.


\section{Construction of the phase transition\label{Construction}}

One important step in the understanding of hybrid stars was done by Glendenning \cite{Glen92, GlenBook} by
realizing that a mixed phase (MP) of deconfined quark matter and hadronic matter inside a neutron star
is not strictly excluded from the star as it was believed before \cite{BaymChin76, ChapNaue76, 
KeisKiss76, FreeMcLe78a}. 
Such MP might have important consequences for our understanding of the observed 
pulsar glitch phenomena or the cooling behavior of neutron stars \cite{HeisPethStau9394}.
It is supposed to form a crystalline lattice of various geometries 
of the rarer phase immersed in the dominant one and probably exists over a wide
range of densities \cite{Glen92, HeisPethStau9394}.
The former predicted exclusion of a MP was owing to the ``freeze out'' of a degree of freedom, 
e.g.~by means of the requirement of charge neutrality in both individually phases. 
This requirement restricts a degree of freedom which the system should originally exploit in its MP
by rearranging electric charge between the two phases.
Every pure phase and a possible MP must of course be charge neutral but in the case of a MP 
this is only a {\em global} charge neutrality. Hence, the quark and hadron components of the MP should 
still have the degree of freedom to reach their energetically favored -- and possibly charged -- state 
if only their volume proportion $\chi$ is always choosen to fulfill the condition of global charge 
neutrality
\begin{equation} \label{globalcharge}
\chi \, q_{QP} + (1-\chi) \, q_{HP} = 0.
\end{equation}
Here $q_{QP}$ and $q_{HP}$ denote the charge density of the quark phase (QP) and the hadronic phase (HP).
The volume proportion $\chi$ of the phases denotes the volume fraction
\begin{equation} 
\chi = \frac{V_{QP}}{V_{QP}+V_{HP}}
\end{equation}
occupied by the QP.
If we do not allow the system to exploit this degree
of freedom we will end up with a MP of constant pressure as it is e.g.~known from a gas-liquid phase 
transition in a familiar one-component system like water \cite{Glen92, GlenBook}. Since the pressure inside 
a star must fall monotonically from its interior to its surface, a constant pressure MP could 
not exist over a finite region inside the star. 
A construction of the first-order phase transition in the sense of Glendenning leads to quite different results. 
As we will see below, the pressure varies continuously in the MP with the volume proportion $\chi$  
from $\chi=0$ (pure HP) to $\chi=1$ (pure QP) leading to a MP of finite extent inside the star \cite{Glen92, GlenBook}.

Following Glendenning, we now write down the Gibbs condition for the equilibrium between QP and HP.
To describe the QP including $u$, $d$, $s$-quarks and electrons, we initially have to deal with four chemical
potentials ($\mu_u$, $\mu_d$, $\mu_s$ and $\mu_e$). Since we require that the phase is in $\beta$-equilibrium, 
the weak reactions
\begin{eqnarray}
d & \longrightarrow & u+e^- +\bar{\nu}_{e^-}, \\
s & \longrightarrow & u+e^- +\bar{\nu}_{e^-}, \nonumber \\
s + u & \longleftrightarrow & d + u, \nonumber
\end{eqnarray}
imply
\begin{eqnarray}
\mu_d & = & \mu_u+\mu_e, \\
\mu_s & = & \mu_u+\mu_e \nonumber
\end{eqnarray}
and reduce the four chemical potentials to two independent ones. 
The chemical potentials of the neutrinos can be taken to be zero because they can diffuse out of the star.
For the remaining two chemical potentials we choose the pair ($\mu_n$, $\mu_e$) where
$\mu_n$ is the neutron chemical potential which is related to the quark chemical potentials through the 
linear combination $\mu_n \equiv \mu_u+2 \mu_d$. 
The same reduction is possible for the HP where we also end up with ($\mu_n$, $\mu_e$). 
%
%
\begin{figure}[ht]
\[\mbox{\tt figure2.gif}\]
\caption{Pressure surfaces, HP = $p_{HP}(\mu_n, \mu_e)$, QP = $p_{QP}(\mu_n, \mu_e)$, 
A = pure charge neutral hadron phase ($p_{HP}>p_{QP}$), B = pure charge neutral quark phase ($p_{QP}>p_{HP}$), 
MP = mixed phase ($p_{QP}=p_{HP}$),  
\mbox{$B^{1/4}=165$ MeV}, \mbox{$g=3$, $K=300$ MeV.}}
\label{phasetransition}
\end{figure}
%
At temperature $T=0$ the Gibbs condition for mechanical and chemical equilibrium between the both phases 
now reads
\begin{equation} \label{pMP}
p_{HP}(\mu_n, \mu_e) = p_{QP}(\mu_n, \mu_e).
\end{equation}
The pressure of both phases spans up a two dimensional surface over the ($\mu_n$, $\mu_e$) plane. 
This situation is shown in fig.~\ref{phasetransition} for a specific choice of parameters . 
The two surfaces HP and QP possess an intersection curve MP corresponding 
to all possible equilibrium states of the mixed phase where $p_{HP}=p_{QP}$ holds. 
For every point on this curve we can calculate $\chi$ by use of (\ref{globalcharge})  
to fulfill global charge neutrality of the mixed phase. From this, the corresponding energy density $\epsilon$ 
can be calculated by
\begin{equation}
\epsilon = \chi \, \epsilon_{QP} + (1-\chi) \, \epsilon_{HP}.
\end{equation}
Hence, by starting with a point on the MP curve -- which corresponds through (\ref{pMP}) to a given pressure $p$ -- 
we can calculate
the energy density $\epsilon$ to finally obtain the EOS of the MP in the form $p=p(\epsilon)$. 
The EOS for the pure HP and pure QP was already discussed in the previous sections. 
The pure HP is realized at a given $(\mu_n, \mu_e)$ if $p_{HP} > p_{QP}$ while the pure QP is realized if 
$p_{HP} < p_{QP}$ holds. 
The pure charge neutral phases are given by curve A for the HP at low pressure 
and by B for the QP in the high pressure regime (see fig.~\ref{phasetransition}). These curves are defined by the
constraint of charge neutrality ($q_{HP}(\mu_n, \mu_e)=0$ and $q_{QP}(\mu_n, \mu_e)=0$ for curve A and B, respectively).
The resulting pressure curve (which finally corresponds to the complete EOS for hybrid stars) is therefore
given by the white curve A-MP-B. 
Obviously the pressure varies continuously along the MP curve and is monotonically increasing.
This clearly shows that a constant pressure construction is not even sufficient in the sense of an approximation. 
We end up with an EOS for hybrid stars which suggests the existence of a mixed phase over a finite range inside the star.

\section{The gross structure of neutron stars} \label{GrossStructure}

To investigate first the general influence of medium effects on the mass-radius (MR) relations and the
structure of the stars, 
we consider $g$ in the next subsection as a constant  
parameter ranging from \mbox{$g=0$} (no medium effects) to $g=4$. 
In the following subsection we will restrict the parameter range by using a density dependent running coupling.
To calculate the MR relation of the star we have to integrate the general relativistic equation of hydrostatic 
equilibrium for a non-rotating spherical star, the Tolman-Oppenheimer-Volkoff equation (TOV) \cite{OppeVolk39}.
The EOS enters into the TOV equation solely in the form $p=p(\epsilon)$.


\subsection{Results with fixed coupling constants\label{gconst}}

For constant $g$ the hybrid star EOS obtained from the phase transition construction (sect.~\ref{Construction}) 
in the $p=p(\epsilon)$ form 
is shown in fig.~\ref{eos}a for low and fig.~\ref{eos}b for high energy densities.
%
\begin{figure}[ht]
\[\mbox{\tt figure3.gif}\]
\caption{EOS for various coupling constants $g$ at low densities (a) and high densities (b). 
Shaded regions (MP) correspond to the mixed phase parts of the EOS, 
$\epsilon_0=140$ MeV/fm$^3$, $B^{1/4}=165$ MeV, $K=300$ MeV.}
\label{eos}
\end{figure}
%
For every $g$, the corresponding EOS is divided into three parts. These are -- from low to high densities --
the pure hadronic phase (HP), the mixed phase (MP) of quark and hadronic matter and the pure quark phase (QP). 
To distinguish the three parts, the region below the $p(\epsilon)$ curve corresponding to the MP is shaded gray.
Again we can see that pressure varies continuously in the MP. As already found by other authors \cite{Glen92, Prak},
the onset of the phase transition to the QP may already occur at densities of the order of the normal nuclear density 
$\epsilon_0=140$ MeV/fm$^3\approx2.5 \times 10^{14}$ g/cm$^3$ (fig.~\ref{eos}a). A larger bag constant $B$ would shift the onset to higher densities 
\cite{Prak}. 


Comparing fig.~\ref{eos}a with fig.~\ref{eos}b, it is important to note, that an increase of $g$ from $g=0$ to $g=3$ 
only slightly shifts the onset of the 
phase transition from $\epsilon\approx\epsilon_0$ to $\epsilon\approx 2\,\epsilon_0$, while the onset of the pure
QP is much more sensitive to a change of $g$. It is shifted from $\epsilon\approx4\,\epsilon_0$ to $\epsilon\approx15\,\epsilon_0$ 
(fig.~\ref{eos}b). This suggests that medium effects mainly influence the transition
point between MP and QP while leaving the transition point between HP and MP more or less unchanged. This finally leads
to an enlarged density range of the MP. 
One may now ask for the reason for this behavior. There are indications
that the small influence of the  $HP \rightarrow MP$ transition density on a change of $g$ is due to the fact that the transition density
(i.e. the density of the onset of quarks) is below the onset of the hyperons in the hadronic phase (which is typically between 
$2\,\epsilon_0$ and $3\,\epsilon_0$ \cite{SchaMish96,GlenBook,Gosh95}). This is obviously fulfilled from $g=0$ to $g=3$ 
in fig.~\ref{eos}a. But the transition density $\epsilon \approx 2\,\epsilon_0$ 
of the $g=3$ result is just at the threshold of the onset of hyperons. Indeed, if we further increase $g$ to say $g=3.5$ 
(which is not shown in fig.~\ref{eos}a), we find that the transition density now strongly increases 
to $\epsilon \approx 12\,\epsilon_0$. 
We can get a qualitative feeling for this behavior if we look again at fig.~\ref{phasetransition} which shows the $g=3$ 
calculation. The point where the pure charge neutral HP curve (A) meets the MP curve corresponds to $\epsilon \approx 2\,\epsilon_0$. 
Up to about this density, the chemical potentials of curve A basically increase in the $\mu_e$ direction which corresponds to a slight 
increase of $\epsilon$. Hence, as long as the transition point for a given $g$ is on this part of the A curve, the transition 
density is rather insensitive on a change of $g$. This is the case for $g\lapprox3$.
At the onset of the hyperons (especially of the negative charged $\Sigma^-$) the electron chemical potential
$\mu_e$ does not increase further and remains at roughly about $\mu_e \approx 170-190$ MeV. Now the chemical potentials 
basically increase in the $\mu_n$ direction corresponding to a strong increase of $\epsilon$. Due to the onset of hyperons the
hadronic EOS is softened which shifts the $HP \rightarrow MP$ transition density to higher values. 
The influence of $g$ on this transition density is now comparable
to the large influence on the $MP \rightarrow QP$ transition density. The latter is large because of an approximately constant $\mu_e$
of about $\mu_e\approx20-30$ MeV (see curve B in fig.~\ref{phasetransition}).
Therefore, one could say that the $HP \rightarrow MP$ transition density is only slightly
influenced by medium effects as long as there are no hyperons present at these densities.


Let us now discuss qualitatively how this interesting behavior of the EOS will influence the structure of the star.
For that reason, let us assume a star with a typical {\em central energy density} $\epsilon_c$ of $\epsilon_c = 10\,\epsilon_0$. At $g=0$ this
corresponds to a star of $M \approx 1.4 M_\odot$. If we look at fig.~\ref{eos}b ($g=0$) we find that $\epsilon_c$ is on
the right hand side of the (dark) gray shaded MP region which ends at $\epsilon \approx 4\,\epsilon_0$. 
Hence, the star should possess a QP core. Following the
star from its center outwards (decreasing $\epsilon$) we obtain a large density region 
(from $\epsilon = 10\,\epsilon_0$ down to $4\,\epsilon_0$) where the star is in its QP. If we now increase the coupling
constant to $g=2$ we see how the density range occupied by the pure QP is narrowed. Thus, the extent of the QP core is reduced. 
At $g=3$ our assumed 
$\epsilon_c$ is not sufficient to bear a pure QP core. The star has a MP core.
This is due to the strong increase of the $MP \rightarrow QP$ 
transition density which shifts the onset of a pure QP to $\epsilon \approx 15\,\epsilon_0$ at $g=3$ (fig.~\ref{eos}b). 
Such large densities can only be achieved in rather soft EOS in which the gravitational force is able to compress matter  
to larger central densities. 
This does not apply to our case of an increased coupling constant. 
If we look at fig.~\ref{eos}a we see that, in fact, the EOS gets stiffer
with increasing $g$ (the pressure $p$ increases at fixed $\epsilon$). This implies that the {\em critical 
central energy density} $\epsilon_{crit}$ (i.e. $\epsilon_c$ at the limiting mass) decreases with $g$.
Therefore, increasing the coupling constant leads to a strong increase of the $MP \rightarrow QP$ transition density
while $\epsilon_{crit}$ (as an upper limit of $\epsilon_c$) is decreasing.
This contrary behavior will -- as we will see -- lead to a surprisingly fast disappearance of a pure QP core with increasing 
$g$. 
A pure hadronic star can only be formed if the $HP \rightarrow MP$ transition density exceeds $\epsilon_c$.
Since for $g\lapprox3$ this transition density only slightly increases with increasing $g$ (fig.~\ref{eos}a) we expect  
a star possessing a MP or a QP core.

Finally, we observe on the right hand side of fig.~\ref{eos}b ($\epsilon\gapprox15\,\epsilon_0$) 
an almost similar EOS of the pure QP for different $g$. 
The pure quark matter EOS in the form  $p=p(\epsilon)$ is therefore only slightly softened with 
increasing medium effects as already found in \cite{Sche97}.

Our qualitative discussion of the EOS and the resulting implications for hybrid stars will now be supported 
by solving the TOV equation.
%
\begin{figure}[ht]
\centerline{\epsfig{file=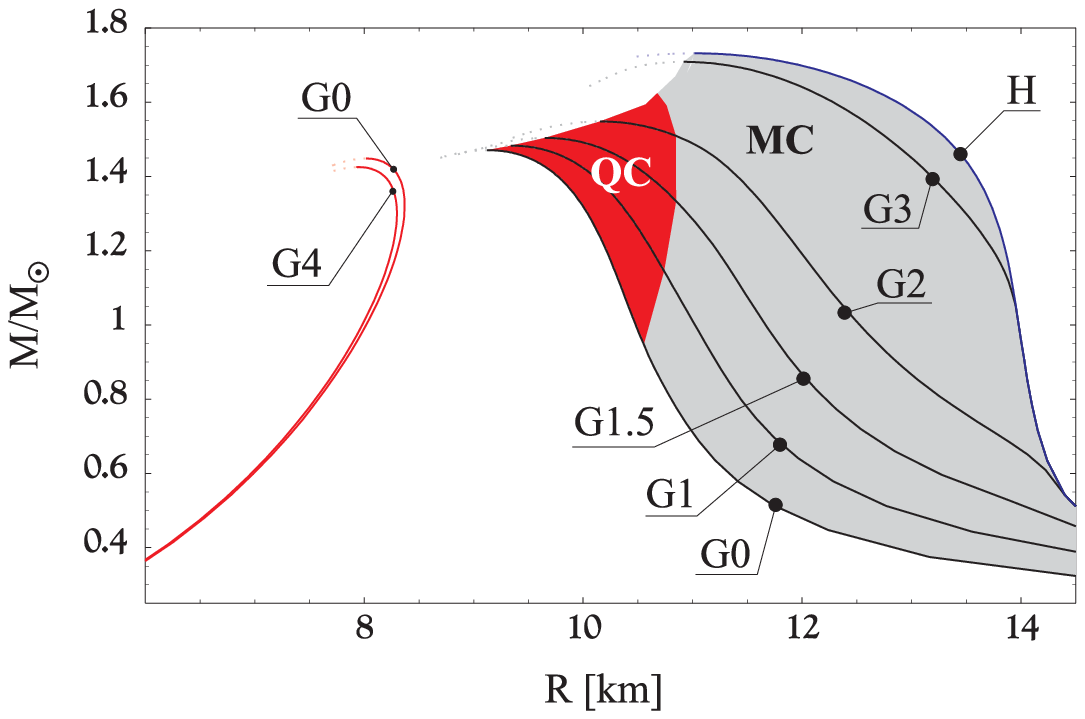,height=8cm}}
\caption{Mass radius relation for pure SQM stars ($R<9$ km) and hybrid stars ($R>9$ km), 
G0 = (g=0), \dots, H = pure hadron, 
QC = star has a quark core, MC = star has a mixed core, $B^{1/4}=165$ MeV, $K=300$ MeV.}
\label{RM1}
\end{figure}
%
Fig.\,\ref{RM1} shows the resulting MR relations (star sequences) for various values of $g$ (e.g.~G1 means $g=1$). 
For comparison, the left hand side ($R<9$ km) shows the pure SQM star (strange star) results, where we do not allow the 
quark phase to transform into hadronic matter. In this case, the MR relation shows the typical $M\propto R^3$ behavior of a 
self bound star at low masses where gravity plays no important role \cite{GlenBook}. It is only somewhat modified at larger masses
where gravity terminates the stable range of the MR relation leading to a limiting mass (mass of the most 
massive star of a MR relation). As already found in \cite{Sche97},
we see that medium effects have only a slight influence on the MR relation of a pure SQM star. 
This situation changes drasticly if we look at the hybrid star results on the right hand side of fig.\,\ref{RM1} ($R>9$ km).
With increasing $g$, the MR relation approaches the curve of the pure hadron star (denoted by H). 
Since the EOS gets stiffer with increasing $g$ the EOS can support a larger limiting mass with a larger corresponding radius.
From the two different shaded regions denoted by QC (QP core) and MC (MP core) we can see 
whether the corresponding central energy density $\epsilon_c$ of a point on a particular MR relation is sufficient to bear a QC 
($\epsilon_c$ has to be larger as the $MP \rightarrow QP$ transition density) or a MC.
The compact QC stars are located in the small vicinity of $M\approx1.4\,M_\odot$ and $R\approx10$ km just below the limiting mass.
The top point of the QC region is reached if the (with $g$) decreasing $\epsilon_{crit}$
gets equal to the increasing $MP \rightarrow QP$ transition density. This is the case for $g\approx2.5$.
At higher $g$, the limiting mass of the sequence is reached before the central density is 
sufficient to develop a QC (see e.g. G3).

Let us now follow e.g. the MR relation for $g=2$ (G2) from low to high central energy densities $\epsilon_c$. 
At very low $\epsilon_c$ ($\epsilon_c\lapprox 1.3\,\epsilon_0$, compare fig.~\ref{eos}a) the G2 curve corresponds 
to the H curve ($R\gapprox14.5$ km).
The corresponding star is a complete hadronic star. With increasing $\epsilon_c$ the star develops a MC and the mass 
is increasing while the radius is decreasing. At $\epsilon_c \gapprox 7\,\epsilon_0$ (compare fig.~\ref{eos}b) the 
central density is sufficient
to bear a pure QP in the center of the star and we enter the QC region. Finally, the limiting mass of the G2 curve is reached at 
$M\approx1.55\,M_\odot$. We see that the largest parts of the MR relation correspond to stars possessing a MC. Only a small
range of mass and radii corresponds to a QC star.

To discuss the radii of the stars and the extent of the quark and mixed cores in detail, we assume a canonical mass of $M=1.4\,M_\odot$.
From fig.\,\ref{RM1} we see that with increasing $g$ the radius increases from originally $R\approx10$ km 
to $R\approx13.5$ km. Such star possesses a QC at small radii below $R\approx 11$ km, a MC at larger radii and 
a pure HP at $R\approx 13.5$ km (H). This is due to the fact that with increasing $g$ the EOS gets stiffer 
owing to a gradually growing impact of the hadronic part of the EOS (fig.~\ref{eos}a).
We can see, that inside such a typical star a QC or a MC is able to exist 
over a wide range of coupling constants from $g=0$ to $g\approx3$ ($\alpha_s\approx0.72$). As already discussed, 
this behavior becomes clear from the only slight influence of $g$ on the $HP \rightarrow MP$ transition density.
In all this quantitative discussions we have of course to keep in mind that the particular values of mass and radius 
obviously depend on the particular choice of the hadronic EOS. 
Nevertheless, we expect that by using a softer hadronic EOS (e.g. a compression modulus of $K=220$ MeV) 
a similar influence of medium effects should be found 
if we also soften the quark matter EOS by use of an appropriately changed bag constant. 
Such dependence should be studied elsewhere \cite{ScheWIP}. 

%
\begin{figure}[ht]
\[\mbox{\tt figure5.gif}\]
\caption{Schematic gross structure of a $M=1.4 M_\odot$ star.}
\label{R}
\end{figure}
%
Fig.~\ref{R} shows the schematic view of the canonical $M=1.4\,M_\odot$ star for different increasing $g$. The different
shells correspond to the quark phase (QP), mixed phase (MP) and hadronic phase (HP).  
We find that a small coupling constant of $g=1.5$ ($\alpha_s\approx0.18$) is able to shrink the radius 
of the pure QP core from $R \approx 6$ km (with neglected medium effects, fig.~\ref{R}a) to $R \approx 3$ km 
(fig.~\ref{R}c). Already at $g=2$ ($\alpha_s\approx0.32$) the pure QP core is vanished completely 
(fig.~\ref{R}d).
Note that in spite of a completely vanishing quark core, the pure hadron phase 
has grown only moderately. One could say that in a wide range of $g$ ($g\lapprox3$) medium effects are not able to
displace the QP in favor of a pure HP. The essential effect is the transformation of the
pure QP into a MP of quark and hadronic matter which therefore dominates the interior of the star.
This behavior can be understood from the strongly increasing $MP \rightarrow QP$ transition density and the only 
slight impact of medium effects on the $HP \rightarrow MP$ transition density.
Only for $g\gapprox3.5$ the phase transition to the QP is completely suppressed (fig.~\ref{R}f).

\subsection{Results with running coupling constant\label{grunning}}

In this section we want to use a running coupling constant in the EOS of SQM instead of assuming $g=$const. as 
we have done in the previous sections. For this purpose we choose the phenomenological running coupling expression
\begin{equation} \label{RC}
g(Q) = \frac{4 \pi}{3} \sqrt{\frac{1}{\ln{Q^2/\Lambda^2}}+\frac{1}{1-Q^2/\Lambda^2}}
\end{equation}
obtained by Shirkov and Solovtsov \cite{ShirSolo97}, where $Q$ is the momentum transfer and $\Lambda$ the QCD
scale parameter (number of quark flavors $n_F=3$). This expression obeys the property of a universal 
limiting coupling of \mbox{$g(0)=4\pi/3$},
while the familiar perturbative expressions possess a singularity at $Q=\Lambda$ (Landau pole). Such singularity and the resulting large $g$ 
in the vicinity of $Q=\Lambda$ is not applicable together with our perturbatively calculated effective masses
where we want to restrict ourselves to $g\lapprox4$ (see also \cite{PeshScheThom97}). 
Since there are no clear results for the running coupling at finite temperature or density 
we take the zero density result (\ref{RC}) and replace $Q$ by the average three-momentum in quark matter $Q=3\mu/4$ 
for every quark chemical potential $\mu$. The quark chemical potential is typically in the range of $\mu \approx 300-500$ MeV
which corresponds to $g\approx2.6-2.3$ for $\Lambda=100$ MeV and $g\approx3.1-2.9$ for $\Lambda=300$ MeV.
We end up with a density dependent running coupling $g(\mu)$ reflecting the behavior of asymptotic freedom.  

%
\begin{figure}[ht]
\centerline{\epsfig{file=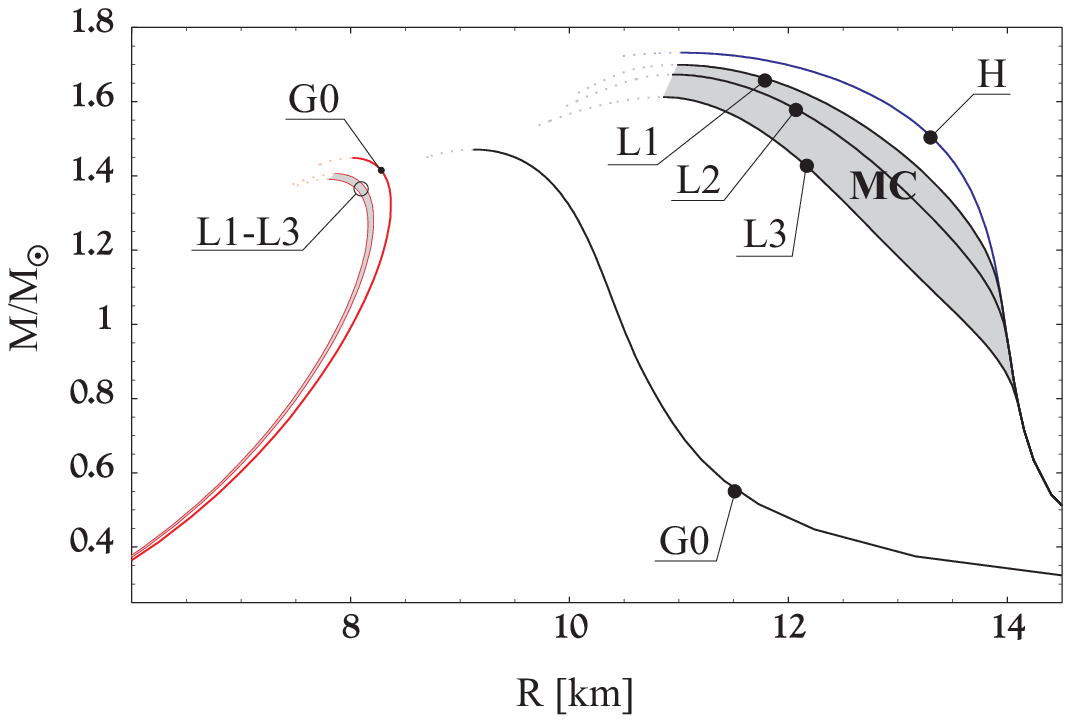,height=8cm}}
\caption{Mass radius relation for pure SQM stars ($R<9$ km) and hybrid stars ($R>9$ km) using a running coupling
constant, 
G0 = (g=0, no medium effects)$, L1 = (\Lambda=100$ MeV), \dots, H = pure hadron, 
MC = star has a mixed phase core, $B^{1/4}=165$ MeV, $K=300$ MeV.}
\label{RM2}
\end{figure}
%
Fig.~\ref{RM2} shows the resulting MR relation using the running coupling constant (\ref{RC}) at 
various QCD scale parameters $\Lambda$. Again, for comparison we show the pure QP results (strange star) 
on the left hand side ($R<9$ km) of fig.~\ref{RM2}, where we do not allow the quark phase to
transform into hadronic matter. The MR relation does not seem to be very 
sensitive to a change of $\Lambda$ from $\Lambda=100$\,MeV (L1) to $\Lambda=300$\,MeV (L3). This is depicted by the gray area.
Comparing the $g=0$ relation (G0) with the running coupling results (L1-L3) we see that 
there is again only a slight influence of medium effects on a pure QP star \cite{Sche97}. 
It is nevertheless interesting to note, that the running coupling leads to a slightly smaller
critical mass compared to the $g=4$ result in fig.~\ref{RM1}. This holds despite the fact that $g(\mu)$ does 
never exceed values of $g\approx 3$ and should therefore naively lead to higher masses.
The right hand side of fig.~\ref{RM2} ($R>9$ km) shows the corresponding hybrid star results. Again, we show 
the MR relation with neglected medium effects (G0) and the relation for the pure hadron star (H). The running coupling 
results are located in a relatively small band just below the pure hadronic solution. 
At least at our particular choice of model parameters, it should be difficult to distinguish between hybrid stars and 
pure hadronic stars by observation of the MR-characteristics of neutron stars only. However, this situation can change 
if lower bag constants are taken \cite{ScheWIP}. In this context see also \cite{HabeTita95Bomb97}. 

Comparing fig.~\ref{RM2} with
fig.~\ref{RM1} one sees that the running coupling results are similar to the $g\approx 3$ results. The central densities
of the stars below the critical densities are not sufficient to yield a pure quark matter core. Hence, the interior of the 
stars for $R\lapprox 14$\,km and $\Lambda$ between \mbox{$\Lambda=100$\,MeV} and \mbox{$\Lambda=300$\,MeV} consists of a mixed phase 
core (MC, shaded region). This is depicted for a canonical \mbox{$M=1.4 M_\odot$} star
in fig.~\ref{R2}. One sees, that the structure of the stars is only slightly sensitive to a change of $\Lambda$. For
\mbox{$\Lambda=200$\,MeV} (fig.~\ref{R2}b) we find a star ($R \approx 13$ km) possessing a mixed core radius of 
\mbox{$R_{MC} \approx 8$ km}. In this case, at least half of the mass of the star is found in its mixed phase. 
%
%
\begin{figure}[hbt]
\[\mbox{\tt figure7.gif}\]
\caption{Schematic gross structure of a $M=1.4 M_\odot$ star using a running coupling constant.}
\label{R2}
\end{figure}


\section{Conclusion \label{conclusion}}
We have investigated the gross structure of non-rotating pure strange quark matter and hybrid stars using the effective 
mass bag model \cite{Sche97} for the description of the quark matter EOS. 
This model is based on the quasi-particle picture where
the quarks acquire medium-dependent effective quark masses generated by the interaction of the quarks with the other
quarks of the system.
We found that medium effects described by this model and parameterized by the
strong coupling constant $g$ reduce the extent of a pure quark matter phase in the interior of a hybrid star 
significantly in favor of a mixed phase. 
For example, the radius of the quark matter core (without medium effects) of originally \mbox{$R\approx 6$ km} 
(see fig.~\ref{R}a) is halved at \mbox{$g=1.5$} (\mbox{$\alpha_s\approx 0.18$}, fig.~\ref{R}c) and is vanished completely
at $g=2$ (\mbox{$\alpha_s\approx 0.32$}, fig.~\ref{R}d). At the same time, the hadronic surface of initial thickness 
\mbox{$D\approx1$ km} is only increased to \mbox{$D\approx2$ km} (fig.~\ref{R}d). So we found that the basic influence of
medium effects on the gross structure of hybrid stars is the reduction of a pure quark matter phase 
in favor of a mixed phase. For a wide range of the coupling constant ($g\lapprox3$, $\alpha_s\lapprox0.72$) 
quark matter is present in the dense 
interior of the star at least as a mixed phase of quark and hadronic matter. These results are confirmed by the 
investigation taking into account a running coupling constant. There we also end up with a hybrid star 
dominated by its mixed phase (fig.~\ref{R2}). Our findings support the importance of the investigation of the 
complex structure of mixed phases.

\bigskip
{\bf Acknowledgements}

One of us (PKS) would like to acknowledge the support from the Alexander-von-Humboldt foundation.
%
%


\begin{thebibliography}{99}

\bibitem{Glen8287}N.K. Glendenning, Phys. Lett. B114 (1982 392; \\
N.K. Glendenning, Z. Phys. A327 (1987) 295.

\bibitem{SchaMish96}J. Schaffner, I.N. Mishustin, Phys. Rev. C53 (1996) 1416.

\bibitem{LiLeeBrow97}G.Q. Li, C.H. Lee, G.E. Brown, Phys. Rev. Lett. 79 (1997) 5214; \\
G.Q. Li, C.H. Lee, G.E. Brown, Nucl. Phys. A625 (1997) 372. \\

\bibitem{SchaGlen98}J. Schaffner-Bielich, N.K. Glendenning, Talk
given at 26th International Workshop on Gross Properties of Nuclei and Nuclear Excitation:
Nuclear Astrophysics (Hirschegg 98), Hirschegg, Austria, 11-17 Jan 1998, nucl-th/9802030. 

\bibitem{BaymChin76}G. Baym, S. Chin, Phys. Lett. B62 (1976) 241.

\bibitem{ChapNaue76}G. Chapline, M. Nauenberg, Nature 264 (1976) 235.

\bibitem{KeisKiss76}B.D. Keister, L.S. Kisslinger, Phys. Lett. B64 (1976) 117.

\bibitem{FreeMcLe78a}B. Freedman, L. McLerran, Phys. Rev. D17 (1978) 1109.

\bibitem{Glen92}N.K. Glendenning, Phys. Rev. D46 (1992) 1274.

\bibitem{HeisPethStau9394}
H. Heiselberg, C.J. Pethick, E.F. Staubo, Phys. Rev. Lett. 70 (1993) 1355; \\
E.F. Staubo, H. Heiselberg, C.J. Pethick, Nucl. Phys. A566 (1994) 577.

\bibitem{GlenPei95}
N.K. Glendenning, S. Pei, Phys. Rev. C52 (1995) 2250.

\bibitem{Prak}
M. Prakash, J.R. Cooke, J.M. Lattimer, Phys. Rev. D52 (1995) 661; \\
M. Prakash, I. Bombaci, M. Prakash, P.J. Ellis, J.M. Lattimer, R. Knorren, Phys.Rept. 280 (1997) 1.

\bibitem{HybRecent}
D. Bandyopadhyay, S. Chakrabarty, S. Pal, Phys. Rev. Lett. 79 (1997) 2176;\\
A. Drago, U. Tambini, astro-ph/9703138.

\bibitem{GlenBook}N.K. Glendenning, Compact Stars (Springer-Verlag, 1997).

\bibitem{GlenPeiWebe97}N.K. Glendenning, S. Pei, F. Weber, Phys. Rev. Lett. 79 (1997) 1603;\\ 
F. Weber, N.K. Glendenning, S. Pei, Invited talk at 3rd International Conference on Physics and 
Astrophysics of Quark Gluon Plasma (ICPAQGP 97), Jaipur, India, 17-21 Mar 1997.


\bibitem{Sche97}K. Schertler, C. Greiner, M.H. Thoma, Nucl. Phys. A616 (1997) 659.
%
\bibitem{Bodm71Witte84}A.R. Bodmer, Phys. Rev. D4 (1971) 1601;\\ 
E. Witten, Phys. Rev. D30 (1984) 272.

%
\bibitem{FahrJaff84}E. Fahri and R.L. Jaffe, Phys. Rev. D30 (1984) 2379.

\bibitem{FreeMcLe78b}B.A. Freedman and L.D. McLerran, Phys. Rev. D16 (1978) 1169.

\bibitem{Lattice}V. Goloviznin and H. Satz, Z. Phys. C57 (1993) 671;\\
A. Peshier, B. K\"ampfer, O.P. Pavlenko, G. Soff, Phys. Lett. B337 (1994) 235; \\
A. Peshier, B. K\"ampfer, O.P. Pavlenko, G. Soff, Phys. Rev. D54 (1996) 2399.

\bibitem{EffeMass}V.V. Klimov, Sov. Phys. JETP 55 (1982) 199; \\ 
H.A. Weldon, Phys. Rev. D26, (1982) 2789; \\
H. Vija and M.H. Thoma, Phys. Lett. B342 (1995) 212; \\
J.-P. Blaizot and J.-Y. Ollitrault, Phys. Rev. D48 (1993) 1390.

\bibitem{GoreYang95}M.I. Gorenstein, S.H. Yang, Phys. Rev. D52 (1995) 5206.

\bibitem{ScheWIP}K. Schertler, C. Greiner, M.H. Thoma, work in progress.

\bibitem{PeshScheThom97}A. Peshier, K. Schertler, M. H. Thoma, 
to be published in Annals of Physics, hep-ph/9708434.

\bibitem{RMF}N.K. Glendenning, F. Weber, S.A. Moszkowski, Nucl. Phys. A572 (1994) 693; \\
J.I. Kapusta, K.A. Olive, Phys. Rev. Lett. 64 (1990) 13; \\ 
J. Ellis, J.I. Kapusta, K.A. Olive, Nucl. Phys. B348 (1991) 345.

\bibitem{Gosh95}S.K. Gosh, S.C. Phatak, P.K. Sahu, Z. Phys. A352 (1995) 457.

\bibitem{BPS}G. Baym, C.J. Pethick, P. Sutherland, Astrophys. J. 170 (1971) 299; \\
G. Baym, H.A. Bethe, C.J. Pethick, Nucl. Phys. A175 (1971) 225.

\bibitem{OppeVolk39}J.R. Oppenheimer, G.M. Volkoff, Phys. Rev. 55 (1939) 347.

\bibitem{ShirSolo97}D.V. Shirkov, I.L. Solovtsov, Phys. Rev. Lett. 79 (1997) 1209.

\bibitem{HabeTita95Bomb97}F. Haberl, L. Titarchuk, Astron. Astrophys. 299 (1995) 414; \\
I. Bombaci, Phys. Rev. C55 (1997) 1587.

\end{thebibliography}
\end{document}